# 3D tumor localization through real-time volumetric x-ray imaging for lung cancer radiotherapy

**Ruijiang Li, John H. Lewis, Xun Jia, Xuejun Gu, Michael Folkerts, Chunhua Men, William Y. Song, and Steve B. Jiang[a]**

*Center for Advanced Radiotherapy Technologies and Department of Radiation Oncology, University of California San Diego, La Jolla, CA 92037*

## Abstract

**Purpose**: To evaluate an algorithm for real-time 3D tumor localization from a single x-ray projection image for lung cancer radiotherapy.

**Methods**: Recently we have developed an algorithm for reconstructing volumetric images and extracting 3D tumor motion information from a single x-ray projection (Li *et al* 2010 *Med Phys* **37** 2822-6). We have demonstrated its feasibility using a digital respiratory phantom with regular breathing patterns. In this work, we present a detailed description and a comprehensive evaluation of the improved algorithm. The algorithm was improved by incorporating respiratory motion prediction. The accuracy and efficiency of using this algorithm for 3D tumor localization were then evaluated on 1) a digital respiratory phantom, 2) a physical respiratory phantom, and 3) five lung cancer patients. These evaluation cases include both regular and irregular breathing patterns that are different from the training dataset.

**Results**: For the digital respiratory phantom with regular and irregular breathing, the average 3D tumor localization error is less than 1 mm which does not seem to be affected by amplitude change, period change, or baseline shift. On an NVIDIA Tesla C1060 GPU card, the average computation time for 3D tumor localization from each projection ranges between 0.19 and 0.26 seconds, for both regular and irregular breathing, which is about a 10% improvement over previously reported results. For the physical respiratory phantom, an average tumor localization error below 1 mm was achieved with an average computation time of 0.13 and 0.16 seconds on the same GPU card, for regular and irregular breathing, respectively. For the five lung cancer patients, the average tumor localization error is below 2 mm in both the axial and tangential directions. The average computation time on the same GPU card ranges between 0.26 and 0.34 seconds.

**Conclusions**: Through a comprehensive evaluation of our algorithm, we have established its accuracy in 3D tumor localization to be on the order of 1 mm on average and 2 mm at 95 percentile for both digital and physical phantoms, and within 2 mm on average and 4 mm at 95 percentile for lung cancer patients. The results also indicate that the accuracy is not affected by the breathing pattern, be it regular or irregular. High computational efficiency can be achieved on GPU, requiring 0.1 - 0.3 s for each x-ray projection.

## I. INTRODUCTION

In modern lung cancer radiotherapy, it is important to have a precise knowledge of real-time lung tumor position during the treatment delivery[1]. Lung tumor localization methods can be broadly categorized as direct tumor localization, localization via breathing surrogates, and localization of implanted markers. Localization based on implanted markers has been shown to be highly accurate[2]. The major drawback is the invasive procedure of marker implantation and the associated risk of pneumothorax[3]. Surrogate-based localization sometimes involves no additional radiation dose to the patient (*e.g.*, optical localization) but suffers from the varying relationship between tumor motion and surrogates both intra- and inter- fractionally[4, 5]. Current radiotherapy treatment machines are usually equipped with an on-board imaging system, which consists of one kV x-ray source and one flat panel imager. Lung tumors may thus be tracked directly without implanted markers in fluoroscopic images acquired by the on-board imaging system[6-9]. Drawbacks of existing direct localization techniques include: 1) they work well mainly for images acquired from the anterior-posterior (AP) direction and the tumors with well-defined boundaries; 2) they usually require the training fluoroscopic data acquired and labeled prior to treatment; 3) the third dimension of tumor motion perpendicular to the x-ray imager cannot be resolved.

These drawbacks may be overcome if a good lung motion model is used for direct tumor localization. Zeng *et al.* used a generic B-spline lung motion model[10] to estimate 3D respiratory motion from cone beam projections. However, because of the large number of parameters in the model, multiple projections (over a 180° gantry rotation) are required to obtain a reasonable estimate for the model parameters. The estimation is thus retrospective and the computation takes quite a long time (several hours on MATLAB). It is clearly not sufficient for use in real-time tumor localization, which has to be done on a sub-second scale. Zhang *et al.* developed a lung motion model based on principal component analysis (PCA), which can efficiently represent the lung motion with only a few eigenvectors and PCA coefficients[11]. The PCA lung motion model has recently been shown to bear a close relationship with the physiological 5D lung motion model proposed by Low *et al.*[12] on a theoretical basis[13]. It is the implicit regularization and high efficiency of the PCA lung motion model that allows one to derive the dynamic lung motion in a reasonably accurate and efficient way, given a very limited amount of information, *e.g.*, a single x-ray projection.

In our previous work[14], we have described an algorithm based on the PCA lung motion model and demonstrated the feasibility of reconstructing volumetric images and extracting 3D tumor motion information in real time from a single x-ray projection using a digital respiratory phantom with regular breathing patterns. In this paper, we will present a detailed description of our improved algorithm and extend the scope of our previous work. In particular, the new contributions of this work relative to our previous work[14] are: 1) a comprehensive evaluation of the algorithm for 3D tumor localization using digital and physical respiratory phantoms as well as lung cancer patients with both regular and irregular breathing patterns, which are different from those in the training dataset; 2) improved efficiency obtained from better initialization of the PCA coefficients using prediction methods.

## II. METHODS AND MATERIALS

For completeness, we first briefly describe the methods for volumetric image reconstruction, which were first presented by Li *et al*[14]. We also discuss ways to further speed up the optimization. Then, we test our algorithm using cone beam CT (CBCT) projections in three different scenarios, namely, a digital respiratory phantom, a physical respiratory phantom, and five lung cancer patients. We reconstructed the volumetric image and derived the 3D tumor

98  location corresponding to each CBCT projection. To quantify the accuracy of our algorithm for
99  tumor localization, we used the average absolute error (denoted as $\bar{e}$) as well as the absolute error
100  at 95 percentile (denoted as $e_{95}$).
101
102  **II.A. Methods**
103
104  Our method for volumetric image reconstruction and 3D tumor localization uses 4DCT from
105  treatment simulation as a priori knowledge. Deformable image registration is performed between
106  a reference phase and the other phases, resulting in a set of deformation vector fields (DVFs).
107  This set of DVFs can be represented efficiently by a few eigenvectors and coefficients obtained
108  from PCA. By changing the PCA coefficients, we can generate a new DVF. A new volumetric
109  image can be obtained when the DVF is applied on the reference image. We then optimize the
110  PCA coefficients such that its computed projection image matches the measured one of the new
111  volumetric image. The 3D location of the tumor can be derived by applying the inverted DVF on
112  its position in the reference image.
113
114  In the PCA lung motion model[11], the DVF relative to a reference image as a function of space and
115  time is approximated by a linear combination of the sample mean vector and a few eigenvectors
116  corresponding to the largest eigenvalues, *i.e.*,

117  $$\mathbf{x}(t) \approx \bar{\mathbf{x}} + \sum_{k=1}^{K} \mathbf{u}_k w_k(t) . \tag{1}$$

118  where $\mathbf{x}(t)$ is the parameterized DVF as a function a space and time, $\bar{\mathbf{x}}$ is the mean DVF with
119  respect to time; $\mathbf{u}_k$ are the eigenvectors obtained from PCA and are functions of space; the
120  scalars $w_k(t)$ are PCA coefficients and are functions of time. It is worth mentioning that the PCA
121  motion model is so flexible that it is capable of represent tumor motion which is beyond that of
122  the training set. For instance, if the PCA coefficients are allowed to change arbitrarily, then given
123  two eigenvectors, the tumor can move anywhere in a plane defined by the corresponding entries
124  in the eigenvectors, and thus hysteresis motion can be handled; given three eigenvectors, the
125  tumor can move anywhere in the space. The reason for this is that each voxel is associated with
126  three entries (corresponding to three canonical directions in space) in each and every eigenvector.
127  Therefore, each eigenvector defines a vector (or a direction) in space along which each voxel
128  moves; two eigenvectors span a plane; and three eigenvectors span the entire 3D space where the
129  tumor can move. This important feature enables the PCA motion model to capture a wide range
130  of tumor motion trajectories. In this study, we kept the first three PCA coefficients in the PCA
131  lung motion model. This is mainly motivated by the fact that three is the least number of PCA
132  coefficients that are capable of capturing a full 3D tumor motion trajectory. On the other hand,
133  using more PCA coefficients, although more flexible, could lead to the overfitting problem and is
134  not necessary.
135
136  After we have obtained a parameterized PCA lung motion model, we seek a set of optimal PCA
137  coefficients such that the projection of the reconstructed volumetric image corresponding to the
138  new DVF matches with the measured x-ray projection. Denote $\mathbf{f}_0$ as the reference image, $\mathbf{f}$ as
139  the reconstructed image, $\mathbf{y}$ as the measured projection image, and $\mathbf{P}$ as a projection matrix
140  which computes the projection image of $\mathbf{f}$ . The cost function is:

141  $$\min J(\mathbf{w}, a, b) = \left\| \mathbf{P} \cdot \mathbf{f}(\mathbf{x}, \mathbf{f}_0) - a \cdot \mathbf{y} - b \cdot \mathbf{1} \right\|_2^2 \tag{2}$$
$$s.t. \quad \mathbf{x} = \bar{\mathbf{x}} + \mathbf{U} \cdot \mathbf{w}$$

142  where, $\mathbf{U}$ is a matrix whose columns are the PCA eigenvectors and $\mathbf{w}$ is a vector comprised of
143  the PCA coefficients to be optimized, $\mathbf{x}$ is the parameterized DVF, and $\|\cdot\|_p$ denotes the standard
144  vector $l_p$-norm. Because the computed and measured projection images may have different pixel

145 intensity levels, we assume there exists a linear relationship between them and introduce two
146 auxiliary parameters $a$ and $b$ to account for the differences. In the Appendix we describe in detail
147 the optimization of the cost function with respect to $\mathbf{w}, a, b$.

148

149 Selecting good initial conditions for the parameters may help speed up the convergence of the
150 algorithm. In our previous work[14], we
151 initialized the PCA coefficients to those
152 obtained for the last projection. A better
153 approach is to consider the last few
154 projections and initialize the PCA
155 coefficients to their predicted values,
156 taking into account the respiratory
157 dynamics. This might become important
158 when the imaging frequency is low. There
159 is a vast amount of literature on respiratory
160 motion prediction models[15-17] that we can
161 use. In this paper, we used a simple linear
162 prediction model[16] which predicts the
163 current sample using a linear combination
164 of several previous samples, *i.e.*,

165
$$w_{k0}(t) = \sum_{l=1}^{L} c_{kl} \cdot w_k^*(t-l) \qquad (3)$$

166 where, $w_{k0}(t)$ is the initial guess for the $k$-
167 th PCA coefficients for the current
168 projection; $w_k^*(t-l)$ is the optimized $k$-th
169 PCA coefficients for previous $l$-th
170 projections. Note that each PCA coefficient
171 has a prediction model associated with it
172 since they have different dynamics.

173

174 In this paper, we utilize the training set
175 (*i.e.*, 4DCT) to build the prediction model
176 (alternatively, the first few breathing cycles
177 during the CBCT scan may be used to
178 build the prediction model). Because the
179 sampling rate for 4DCT and CBCT scan is
180 different (for instance, for a patient with a
181 4-sec breathing period, the sampling rate
182 for 4DCT is 2.5 Hz if 10 phases were
183 reconstructed; on the other hand, the
184 sampling rate for CBCT projections is
185 faster, usually on the order of 10 Hz for a
186 typical 1-min scan with around 600
187 projections), we first interpolated the PCA
188 coefficients from the training set and re-
189 sampled to have the same frequency as in
190 CBCT scan. Then the model fitting process
191 was done separately for each of the PCA
192 coefficients since they are assumed to be
193 independent in the PCA motion model.
194 Here, we would like to point out that the

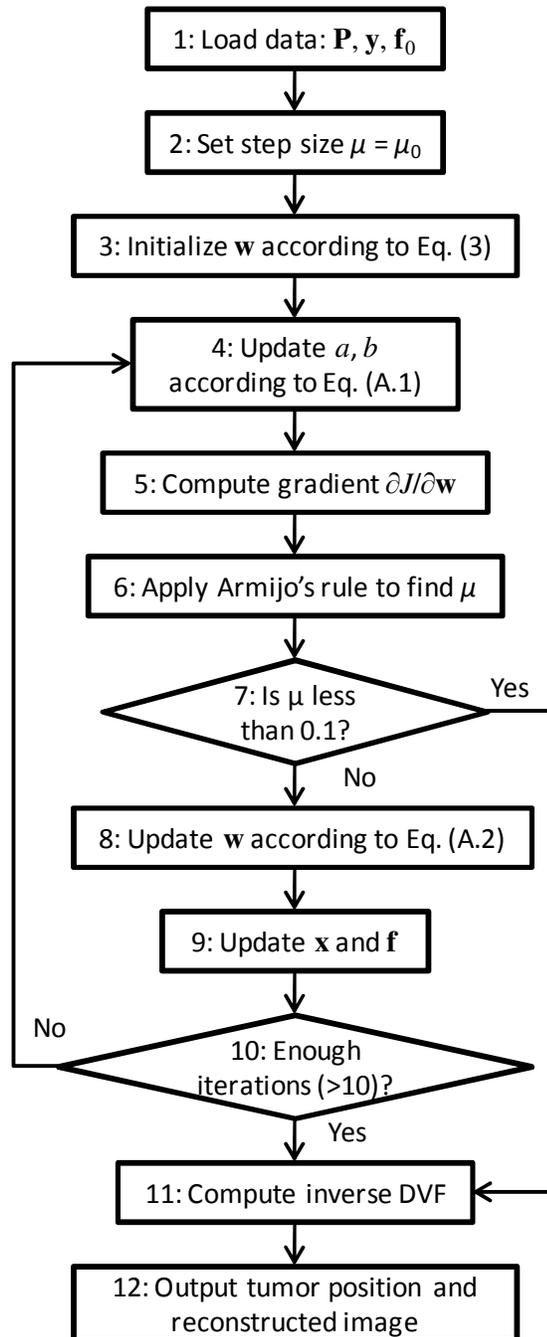

**Fig. 1.** The flow chart of the proposed algorithm for volumetric image reconstruction and tumor localization based on a single x-ray projection. For the meaning of step size and gradient as well as expressions of Eq. (A.1), Eq. (A.2), refer to the Appendix.

195  primary goal of the prediction model is to obtain a "good" initial guess for the PCA coefficients,
196  so that the efficiency of the algorithm may be improved. For this purpose, the linear model used
197  here seems sufficient.
198
199  In order to get the current tumor position, we note that the DVF found by Eq. (2) is a pull-back
200  DVF, which is defined on the coordinate of the new image and tells how each voxel in the new
201  image moves. It cannot be used directly to calculate the tumor position in the new image unless
202  the DVF is rigid everywhere. To get the correct tumor position, we need its inverse, *i.e.*, the push-
203  forward DVF, which is defined on the coordinate of the reference image and tells how each voxel
204  in the reference image moves. In this work, we adopted an efficient fixed-point algorithm for
205  deformation inversion, which has been shown to be about 10 times faster and yet 10 times more
206  accurate than conventional gradient-based iterative algorithms[18]. Since we are primarily
207  concerned with the tumor motion, it is not necessary to apply the deformation inversion
208  procedure on the entire image. Instead, it is applied on only one voxel, which corresponds to the
209  tumor center of volume in the reference image. Finally, we summarize our algorithm in a flow
210  chart, as shown in Fig. 1.
211
212  To achieve real-time efficiency, we have implemented our algorithm on graphic processing units
213  (GPUs). For this work, we used compute unified device architecture (CUDA) as the
214  programming environment and an NVIDIA Tesla C1060 GPU card.
215
216  For evaluation purposes, we used the end of exhale (EOE) phase in 4DCT as the reference image
217  and performed deformable image registration (DIR) between the EOE phase and all other phases.
218  The DIR algorithm used here is a fast demons algorithm implemented on GPU[19]. Then PCA was
219  performed on the nine DVFs from DIR and three PCA coefficients and eigenvectors were kept in
220  the PCA lung motion model (except for the physical phantom, where the motion is along the
221  longitudinal axis so that only one PCA coefficient is necessary). For the prediction model in Eq.
222  (3), we selected $L = 2$ in all the cases.
223

## II.B. Digital respiratory phantom

226  The algorithm was first tested using a non-uniform rational B-spline (NURBS) based cardiac-
227  torso (NCAT) phantom[20]. In this work, we considered three fundamental breathing parameters:
228  namely amplitude, period, and baseline shift relative to the training dataset. In the case of regular
229  breathing, we first used the same breathing parameters as those in the training set and generated
230  testing volumetric images. We then systematically changed the breathing pattern such that each
231  time, only one of the three parameters was changed while the other two were kept the same as
232  those in the training set. In doing so, we can independently evaluate the effects of each of the
233  three breathing parameters. The different breathing parameters considered in this work are:
234  amplitude (diaphragm motion: 1 cm and 3 cm), period (3 s and 5 s), and baseline shift (1 cm
235  toward the EOE phase and EOI phase). In the case of irregular breathing, we used a scaled
236  version of the breathing signal of a real patient acquired by the Real-time Position Management
237  (RPM) system (Varian Medical Systems, Palo Alto, CA, USA). The dynamic NCAT phantom
238  motion can be generated by specifying the diaphragm SI motion. In this case, the scaled RPM
239  signal was then used as a substitute for the diaphragm SI motion of the NCAT phantom. The
240  breathing signal from RPM has a varying breathing amplitude (between 1.0 and 2.5 cm), period
241  (between 3.5 and 6.2 s), and a baseline shift of about 1 cm.
242
243  After we obtained the dynamic NCAT phantom, we then simulated 360 cone beam projections
244  using the Siddon's algorithm[21] from angles that are uniformly distributed over one full gantry
245  rotation (*i.e.*, the angular spacing between consecutive projections is 1°). For instance, the cone
246  beam projection at angle 180° corresponds to the dynamic phantom at 30 sec, and the projection

247     at angle 360° corresponds to the dynamic phantom at 60 sec, and so on. Since the gantry rotation
248     lasts 60 seconds, the sampling rate of the cone beam projections is 6 Hz. We did not add any
249     quantum noise in the simulated x-ray projections for the digital phantom cases. However,
250     intrinsic quantum noise is present in all physical phantom and patient projections and no noise
251     reduction was attempted on those images. The imager has a physical size of 40×30 cm$^2$. For
252     efficiency considerations, we down-sampled every measured projection image to a resolution of
253     200×150 (pixel size: 2×2 mm$^2$).
254
255     **II.C. Physical respiratory phantom**
256
257     The algorithm was also tested on a simple physical respiratory phantom. The phantom consisted
258     of a cork block resting on a platform that can be programmed to undergo translational motion in
259     one dimension. This platform was used to simulate the SI respiratory motion of a lung tumor.
260     Inside the cork block were embedded several tissue-like objects including a 2.5 cm water balloon
261     which was used as the target for localization. More detailed descriptions of the phantom can be
262     found in Lewis *et al.*[9].
263
264     4DCT of the physical phantom was acquired using a GE four-slice LightSpeed CT scanner (GE
265     Medical Systems, Milwaukee, WI, USA) and the RPM system. During the 4DCT scan, the
266     physical phantom moved according to a sinusoidal function along the SI direction, with a peak-
267     to-peak amplitude of 1.0 cm. In order to calculate the correct projection matrix and image for the
268     algorithm, the CT image corresponding to the EOE phase in 4DCT was re-sampled in space to
269     have the same spatial resolution as CBCT and then rigidly registered to the CBCT. Since only
270     rigid registration was performed at this stage, in reality, paired kV radiographs may also be used
271     to achieve this task during patient setup. Cone-beam projections were acquired using Varian On-
272     Board Imager 1.4 in full-fan mode with 100 kVp, 80 mA and 25 ms exposure time. The x-ray
273     imager has a physical size of 40×30 cm$^2$ and each projection image has a resolution of 1024×768.
274
275     For testing purposes, we performed two CBCT scans for the physical phantom: one with regular
276     breathing, and the other with irregular breathing. For the case of regular breathing, the phantom
277     moved according to a sinusoidal function along the SI direction, with an increased peak-to-peak
278     amplitude of 1.5 cm. For the case of irregular breathing, the phantom moved according to the
279     same RPM signal used for the digital phantom as described before, with the peak-to-peak
280     amplitude scaled to 1.5 cm. In both cases, 359 bone beam x-ray projections were acquired over an
281     arc of around 200° with a frequency of about 10.7 Hz. The CBCT scans for the physical phantom
282     lasted for about 36 s, so only the first 36 s of the RPM signal was used during the scan. Since the
283     motion of the physical phantom during the CBCT scan is known, it was used as the ground truth
284     to evaluate the accuracy of target localization.
285
286     The following preprocessing was performed on each of the cone beam x-ray projections. First,
287     due to the presence of the full-fan bow-tie filter near the x-ray source, a systematic bias will be
288     introduced in the linear relations between the image intensities of the simulated and measured
289     projections. So an in-air x-ray projection with exactly the same imaging protocol was acquired in
290     order to correct for the systematic bias. This was done by dividing each pixel intensity value of
291     the projection images acquired for the physical phantom by those in the in-air projection image.
292     Each corrected projection image was then down-sampled by a factor of 4, resulting in a resolution
293     of 256×192 (pixel size: 1.56×1.56 mm$^2$). The rationale and effect of downsampling will be
294     discussed later in the paper.
295
296
297
298

## II.D. Patient data

Finally, the algorithm was evaluated with five patient data sets. Both 4DCT and CBCT were acquired using the same imaging systems as for the physical phantom. The cone beam projections were taken with the imaging system in half-fan mode with 110 kVp, 20 mA and 20 ms exposure time. Similarly to the physical phantom, the CT at the EOE phase was re-sampled in space and then rigidly registered to the CBCT according to the bones. The same preprocessing on each of the raw cone beam projections of the patient was used as for the physical phantom, namely, cutoff, downsampling, and a logarithm transform, except for the bow-tie filter correction, which was based on the half-fan mode. For patient 2, however, since the computed projections of the 4DCT did not have sufficient longitudinal coverage compared with the cone beam projections, an additional 3 cm was cut off from the raw cone beam projections in both superior and inferior directions.

For all patients in this study, there were no implanted fiducial makers and real-time 3D location of the tumor was not available to evaluate our algorithm. Instead, we projected the estimated 3D tumor location onto the 2D imager and compared with that manually defined by a clinician. The patients chosen for this study had tumors that were visible to the clinician in some of the cone beam projections. From the clinician-defined contour, the tumor centroid position was calculated for each projection and used as the ground truth to evaluate the algorithm. We calculated the tumor localization error along the axial and tangential directions, both scaled back to the mean tumor position. The axial direction is defined to be along the axis of rotation on the imager and the tangential direction is perpendicular to the axis of rotation on the imager. For a tumor located near the isocenter, the tumor motion along the axial direction on the imager is roughly a scaled version of its SI motion in the patient coordinate system. On the other hand, the tumor motion along the tangential direction is a mixture of its AP and LR motions depending on the imaging angle. In some special cases, the tumor motion along the tangential direction can be roughly a scaled version of either the AP motion (for imaging angles near LR or RL) or LR motion (for imaging angles near AP or PA). Since the tumor is hard to see in some of the projections, the uncertainty in the definition of ground truth can be large, and may depend on the angle at which a projection is taken. A rough estimate of the uncertainty in the ground truth based on comparing contours drawn from two observers suggests that the error in centroid position is about 1-2 mm on average.

For each patient, approximately 650 projections were acquired over a full gantry rotation with a frequency of about 10.7 Hz. However, since the cone beam scans were performed in the half-fan mode (with the imager shifted about 14.8 cm laterally), and isocenter of the scan was not placed at the tumor, the tumor is only visible in a subset of these projections. For each patient, the tumor was marked by the clinician in the largest continuous set of projections in which the tumor was visible. For the five patients, the number of cone beam projections used for this study ranged between 70 and 281, corresponding to angles of 39º and 155º, respectively.

## II.E. Summary of testing cases

Table 1 summarizes all the testing cases in this work, including 8 digital phantom cases, 2 physical phantom cases, and 5 lung cancer patient cases. The table lists the relevant breathing parameters that were used during the CBCT scan for tumor localization purposes. It also includes information on the tumor size, location, as well as motion characteristics as measured in 4DCT for the 5 lung cancer patients.

**Table 1.** Summary of all testing cases. For the digital phantom, the breathing parameters in the training dataset are: 2 cm amplitude, 4 s period, and 0 cm baseline shift; for cases 2 through 7, all breathing parameters are the same as in training dataset except the one indicated in the table. All parameters for lung cancer patients were measured in 4DCT.

| Case # | Category | Relevant Parameters |
|---|---|---|
| 1 | Digital phantom 1 | Same as training |
| 2 | Digital phantom 2 | Amplitude changed to 1 cm |
| 3 | Digital phantom 3 | Amplitude changed to 3 cm |
| 4 | Digital phantom 4 | Period changed to 3 s |
| 5 | Digital phantom 5 | Period changed to 5 s |
| 6 | Digital phantom 6 | Baseline shift changed to 1 cm toward the EOE phase |
| 7 | Digital phantom 7 | Baseline shift changed to 1 cm toward the EOI phase |
| 8 | Digital phantom 8 | Breathing changed to irregular patient breathing |
| 9 | Physical phantom 1 | Amplitude changed to 1.5 cm (training: 1 cm) |
| 10 | Physical phantom 2 | Breathing changed to irregular patient breathing |
| 11 | Lung cancer patient 1 | 2.7 cm$^3$ tumor at right lower lobe; 20 mm tumor motion |
| 12 | Lung cancer patient 2 | 18.3 cm$^3$ tumor at left lower lobe; 17 mm tumor motion |
| 13 | Lung cancer patient 3 | 8.1 cm$^3$ tumor at left upper lobe; 5 mm tumor motion |
| 14 | Lung cancer patient 4 | 3.3 cm$^3$ tumor at right lower lobe; 11 mm tumor motion |
| 15 | Lung cancer patient 5 | 10.2 cm$^3$ tumor at left lower lobe; 12 mm tumor motion |

# III. Results

**Table 2.** Summary of localization accuracy and computational time for all testing cases. For the description of each case, refer to Table 1. For Cases 1 through 10 (digital and physical phantoms), the localization errors were computed in 3D; for Cases 11 through 15 (lung cancer patients), the localization errors were computed on the axial and tangential directions with respect to the imager.

| Case # | $\bar{e}$ (mm) | | $e_{95}$ (mm) | | Time (s) |
|---|---|---|---|---|---|
| 1 | 0.8 | | 1.8 | | 0.20 ± 0.05 |
| 2 | 0.7 | | 1.7 | | 0.19 ± 0.05 |
| 3 | 0.8 | | 1.8 | | 0.22 ± 0.06 |
| 4 | 0.8 | | 1.8 | | 0.20 ± 0.05 |
| 5 | 0.8 | | 1.8 | | 0.20 ± 0.05 |
| 6 | 0.8 | | 1.6 | | 0.24 ± 0.05 |
| 7 | 0.8 | | 1.8 | | 0.24 ± 0.05 |
| 8 | 0.8 | | 1.6 | | 0.26 ± 0.06 |
| 9 | 0.8 | | 2.2 | | 0.13 ± 0.04 |
| 10 | 0.8 | | 2.4 | | 0.16 ± 0.05 |
| – | Axial | Tangential | Axial | Tangential | – |
| 11 | 1.5 | 1.8 | 3.1 | 3.6 | 0.29 ± 0.07 |
| 12 | 1.9 | 1.8 | 3.7 | 3.8 | 0.31 ± 0.12 |
| 13 | 0.4 | 0.9 | 1.0 | 2.5 | 0.26 ± 0.04 |
| 14 | 0.9 | 1.0 | 2.1 | 2.4 | 0.34 ± 0.09 |
| 15 | 1.1 | 0.4 | 2.7 | 1.2 | 0.33 ± 0.11 |

## III.A. Results for the digital respiratory phantom

Figure 2 shows the SI tumor position estimated by our algorithm as well as the ground truth for the digital respiratory phantom for Cases 1 through 8. The numerical results for localization accuracy and computational time are summarized in Table 2. The average 3D tumor localization error for regular breathing with different breathing parameters is less than 1 mm, which does not seem to be affected by amplitude change, period change, or baseline shift. The same holds true for irregular breathing.

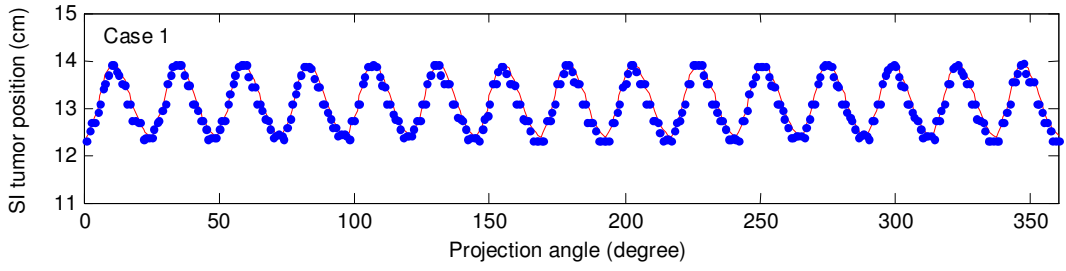

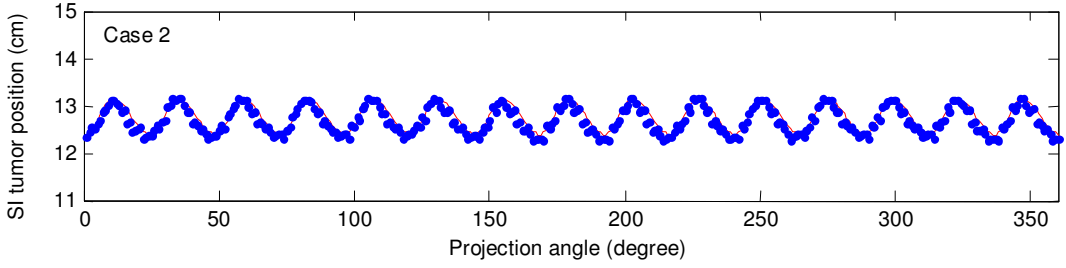

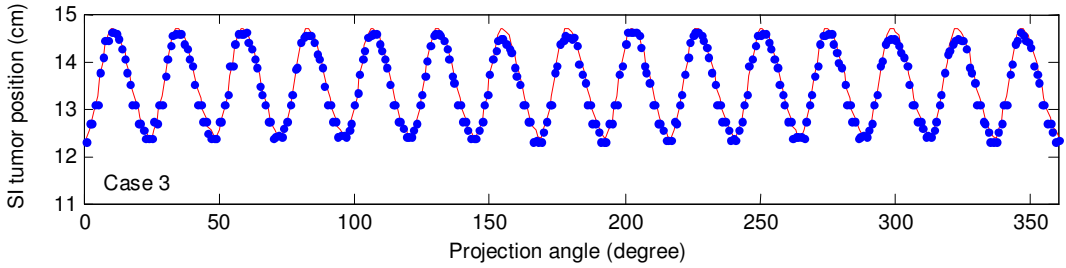

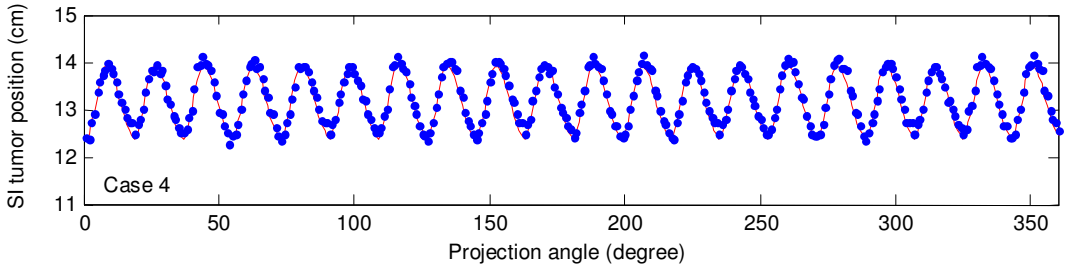

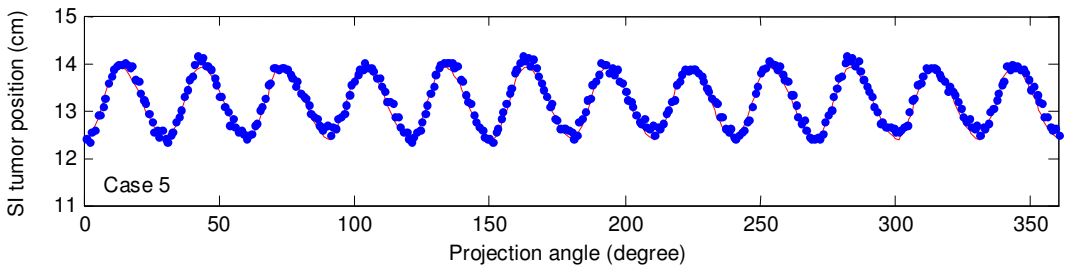

376

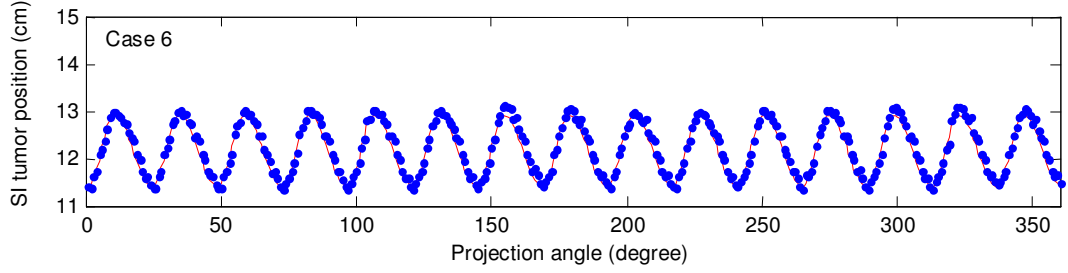

377

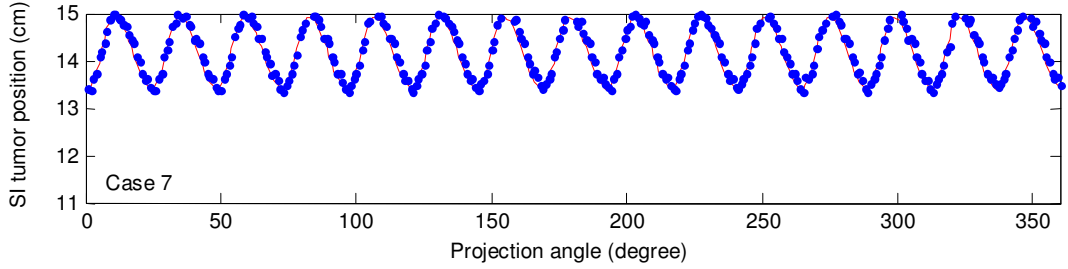

378

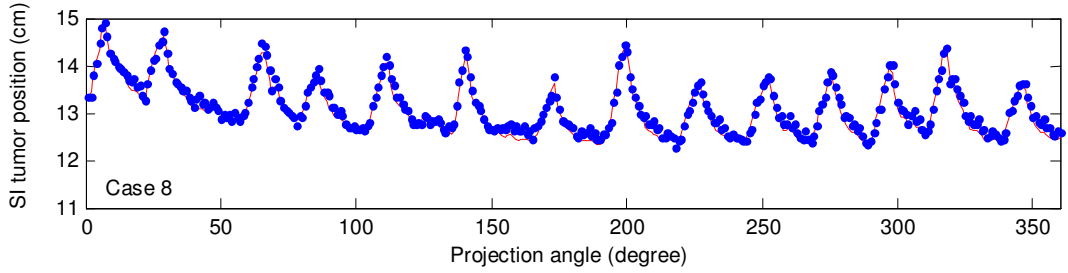

379 **Fig. 2.** SI tumor position estimated by our algorithm (black circles) and the ground truth (solid line) for the
380 digital phantom (Cases 1 through 8).

381

382 From the re-sampled PCA coefficients in the training set, the prediction models were estimated
383 as:

$$w_{1,0}(t) = 1.90w_1^*(t-1) - 0.98w_1^*(t-2)$$

$$w_{2,0}(t) = 1.51w_2^*(t-1) - 0.84w_2^*(t-2)$$ (4)

$$w_{3,0}(t) = 0.97w_3^*(t-1) - 0.84w_3^*(t-2)$$

384

385 By initializing the PCA coefficients to their predicted values using the above equations, the
386 average computation time for both regular and irregular breathing ranges between 0.19 and 0.26
387 seconds using an NVIDIA Tesla C1060 GPU card. Particularly, the average computation time is
388 0.22 seconds in the case of regular breathing with a breathing amplitude of 3 cm. This is about
389 10% less than that previously reported[14] using the PCA coefficients from the previous projection.
390 In the case of irregular breathing, the average computation time is $0.26 \pm 0.06$ seconds. The slight
391 increase in computation time for irregular breathing was mainly due to a decreased accuracy of
392 the prediction model used for parameter initialization.

393

394 **III.B. Results for the physical respiratory phantom**

395

396 Figure 3 shows the estimated and ground truth tumor position in the SI direction for both regular
397 and irregular breathing. The numerical results for localization accuracy and computational time
398 are summarized in Table 2 (Cases 9 and 10). The average 3D tumor localization error for both
399 regular and irregular breathing is less than 1 mm.

Since only one PCA coefficient was used for the physical phantom, the prediction model was estimated as:

$$w_{1,0}(t) = 1.96 w_1^*(t-1) - 0.99 w_1^*(t-2) \tag{5}$$

When PCA coefficients were initialized to their predicted values, the average computation time on GPU is 0.13 seconds and 0.16 seconds, for regular and irregular breathing, respectively. The reduced computation time compared with the NCAT phantom is mainly due to the less number of PCA coefficients used for the physical phantom.

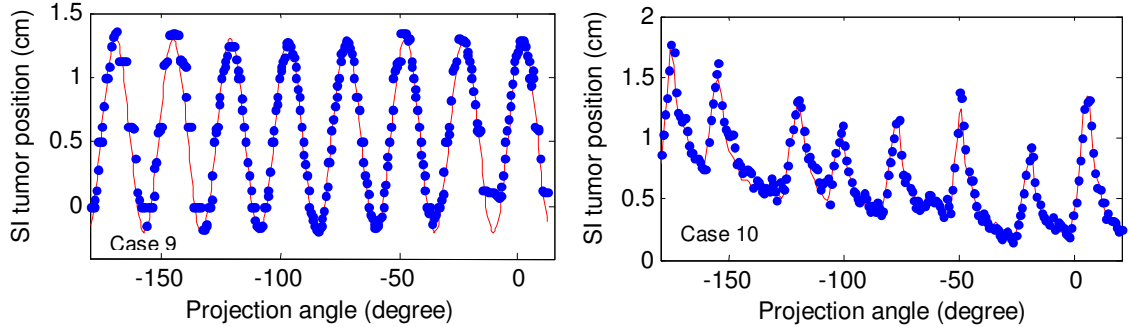

**Fig. 3.** SI tumor position estimated by our algorithm (black circles) and the ground truth (solid line) for the physical phantom (Cases 9 and 10). Left: regular breathing; Right: irregular breathing.

## III.C Results for the patient data

Results for the five lung cancer patients are shown in Fig. 4, where the tumor positions on the axial and tangential directions are shown separately. All five patients had somewhat irregular breathing during the CBCT scans especially for patient 2 (Case 12) and the last breathing cycle of patient 4 (Case 14). Again, the numerical results for localization accuracy and computational time are summarized in Table 2 (Cases 11 through 15). For all five patients, the average 3D tumor localization error is less than 2 mm and the absolute error at 95 percentile is less than 4 mm on both axial and tangential directions. When PCA coefficients were initialized to their predicted values, the average computation time on GPU is between 0.26 seconds and 0.34 seconds for the five patients.

Figure 5 shows the image reconstruction and tumor localization results for patient 1 at an EOE phase and an EOI phase (indicated in Figure 4, Case 11), as well as the corresponding cone beam x-ray projections. Although it is hard to see the tumor on the projections, it is clearly visible in the coronal and sagittal slices of the reconstructed images.

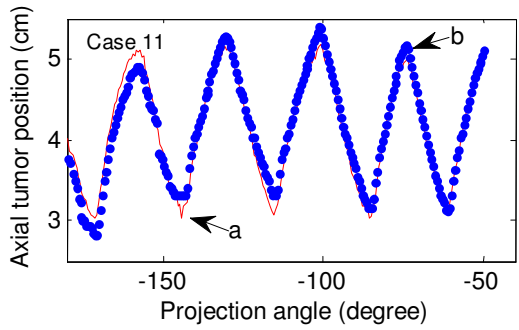
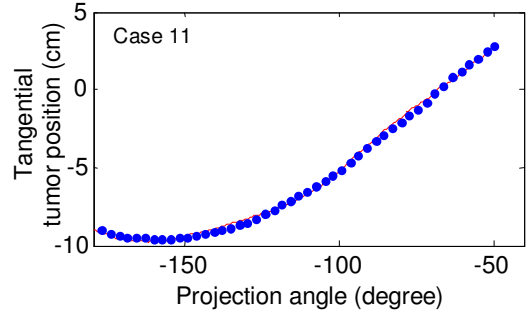

431

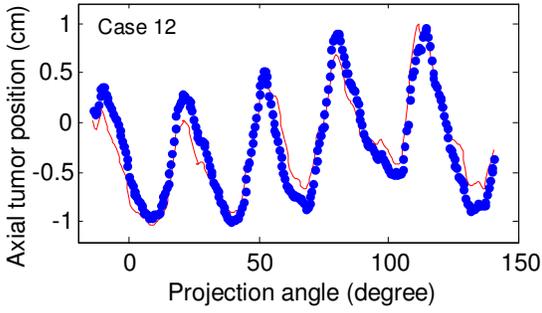
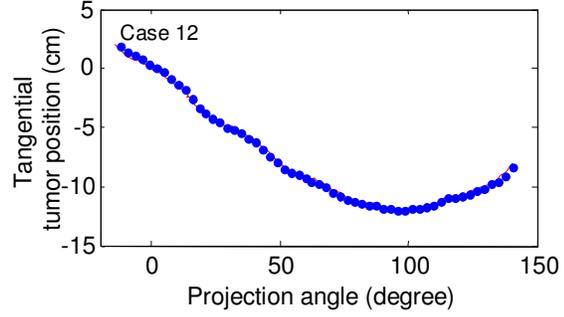

432

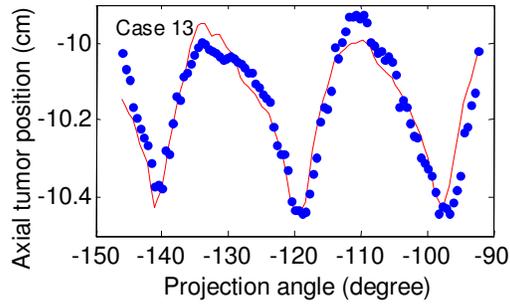
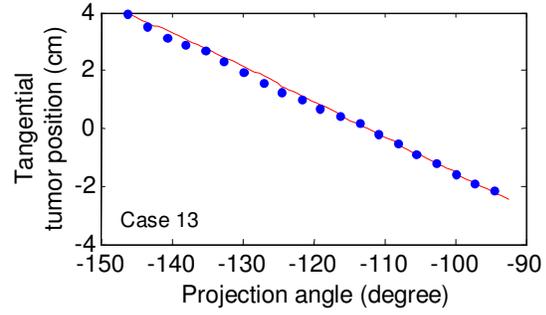

433

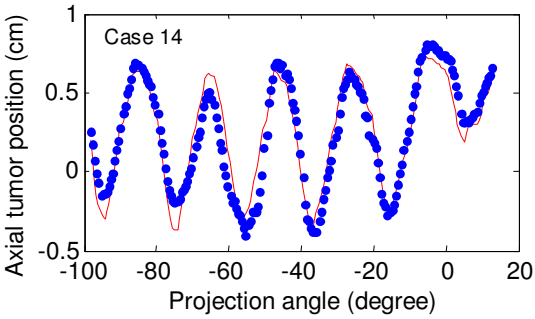
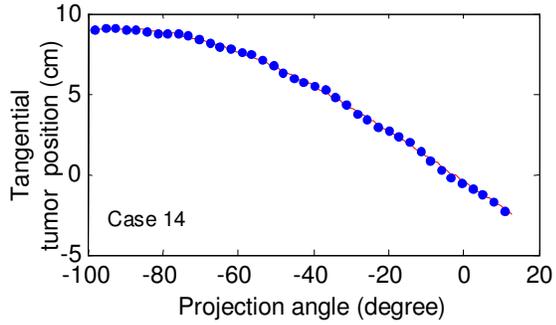

434

435
436
437 **Fig. 4.** Tumor positions estimated by our algorithm (black circles) and the ground truth (solid line) for five
438 lung cancer patients (Cases 11 to 15). Left: axial tumor position; Right: tangential tumor position. The two

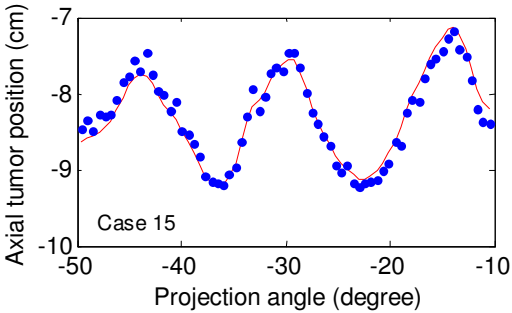
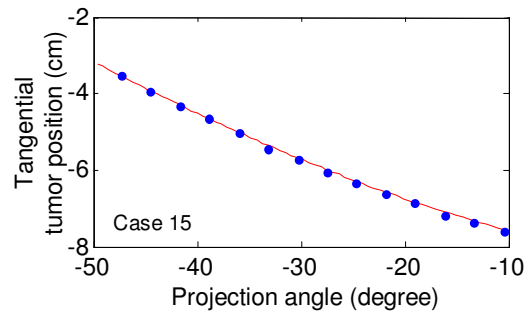

439 arrows ("a" and "b") in the left plot of patient 1 (Case 11) indicate the two angles where the image
440 reconstructions results are shown in Fig. 5.
441

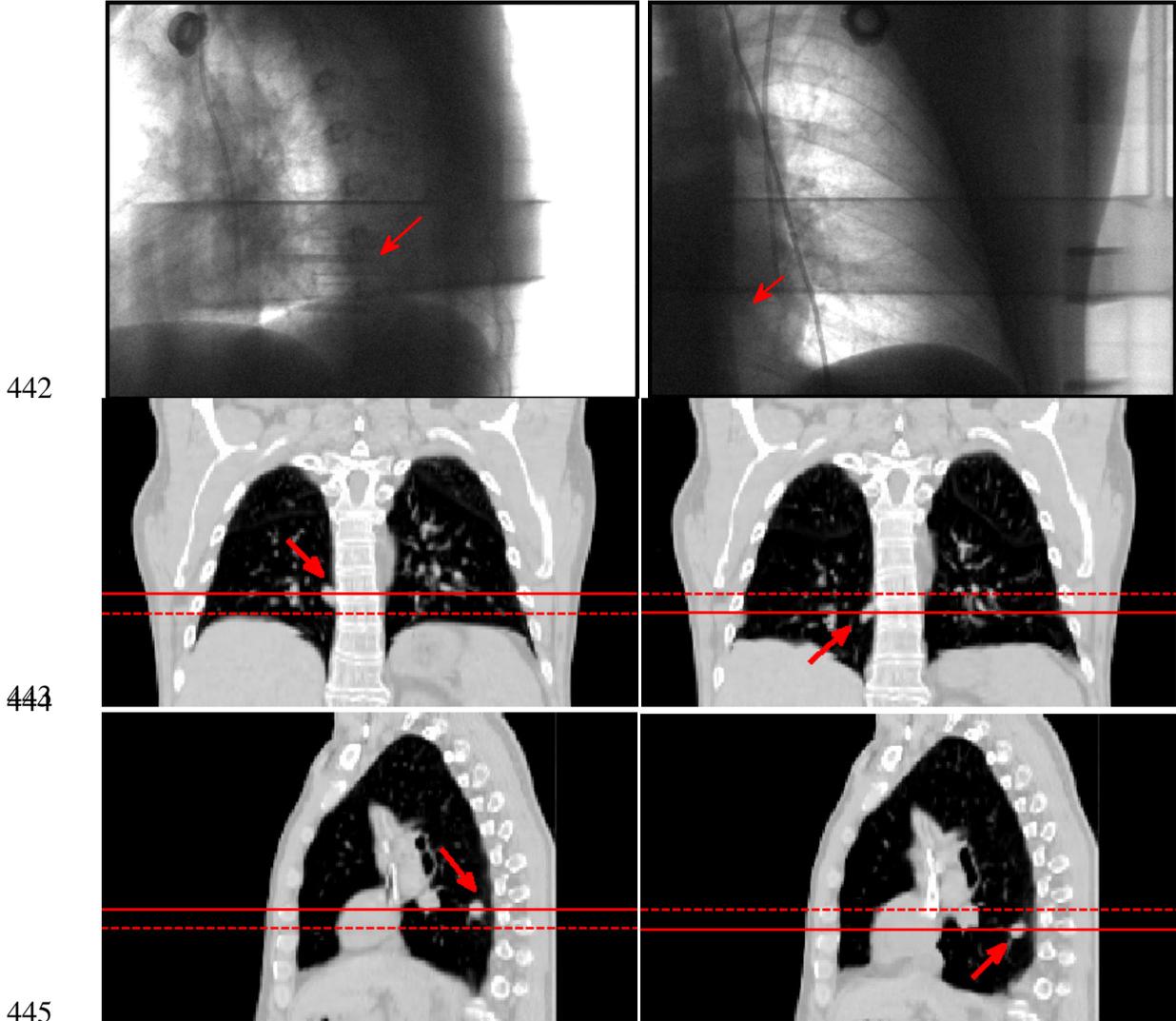

**Fig. 5.** Image reconstruction and tumor localization results for patient 1. Left column: raw cone beam
projection (*top*) and the corresponding coronal and sagittal views of the reconstructed image at angle -
146.5° at an EOE phase (arrow "a" in Fig. 4). Right column: same as left column, except for angle -74.5° at
an EOI phase (arrow "b" in Fig. 4). Red arrow indicates where the tumor was. The solid line represents the
estimated tumor SI position at the current phase; the dashed line represents the estimated tumor SI position
at the other phase.

There is a tradeoff between the localization accuracy and computational time, which is dependent
on the resolution of the projection image. Intuitively, as the image resolution is increased, the
localization error would decrease; at the same time, the resulting computational time will
naturally increase. To investigate the effect of the projection image resolution, we took the
original cone beam x-ray projection images with a resolution of 1024×768 for patient 1 (Case 11)
and down-sample them by a factor of 1, 2, 4, 8, 16, 32 on both dimensions of the projection
images. The resulting images have the resolution of 1024×768, 512×384, 256×192, 128×96,
64×48 and 32×24, respectively. Figure 6 shows the average localization error on tangential and
axial directions as well as computational time versus the downsampling factor. As expected, the
average localization error increases monotonically as the downsampling factor increases.
However, downsampling the projection image by a factor of 4 does not increase the localization error noticeably. On the other hand, the computation time decreases monotonically as the downsampling factor increases, but flattens out once it reaches 8. This supports the choice of a resolution of 256×192 (downsampling factor: 4) for the projection image used in this study, because it achieves a reasonable balance between the localization accuracy and computational time. Whether this observation holds true in general for clinical cases deserves further investigation.

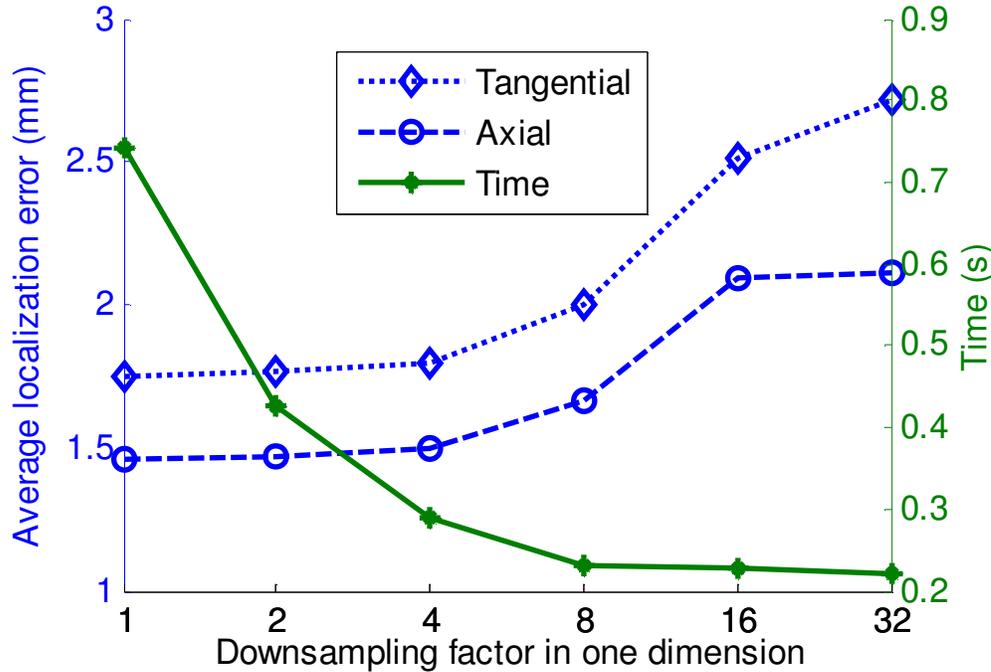

**Fig. 6.** The average localization error on tangential and axial directions as well as computational time versus downsampling factor in one dimension of the x-ray imager for patient 1 (Case 11).

## IV. Discussions

We have developed an algorithm to localize 3D tumor positions in near real-time from a single x-ray image. The algorithm was systematically tested on digital and physical respiratory phantoms as well as lung cancer patients. For the digital phantom, the average 3D tumor localization error is below 1 mm in the case of digital and physical phantoms. For the five lung cancer patients, the average tumor localization error is below 2 mm in both the axial and tangential directions. By utilizing the massive computing power of GPUs, we were able to derive 3D tumor locations from one projection around 0.3 seconds on average. Successful applications of our algorithm in the clinic could lead to accurate 3D tumor tracking from a single x-ray imager. Furthermore, volumetric images can be reconstructed from a single x-ray projection image. This is potentially useful for many clinical applications such as dose verification in lung cancer radiotherapy.

The tumor localization error in this work arises from two different processes: 1), the PCA lung motion model; 2), the 2D to 3D deformable registration between the projection and the reference image. In order to investigate how much each process contributes to the overall tumor localization error, we utilized the ground truth of the 3D lung motion in the digital NCAT phantom and solved for the optimal PCA coefficients by directly minimizing the mean square error between the left and right sides in Eq. (1). This procedure bypassed the 2D to 3D registration process, and thus gives the tumor localization error solely due to the PCA lung motion model. In Case 1, this error is about 0.28 mm over all the projections. Assuming that the above two processes are statistically independent, we readily obtain the error due to the 2D to 3D registration process, which is about 0.75 mm. Therefore, the overall tumor localization error of about 0.8 mm is mainly attributed to the 2D to 3D registration process. In comparison, the error introduced by the PCA lung motion model is insignificant.

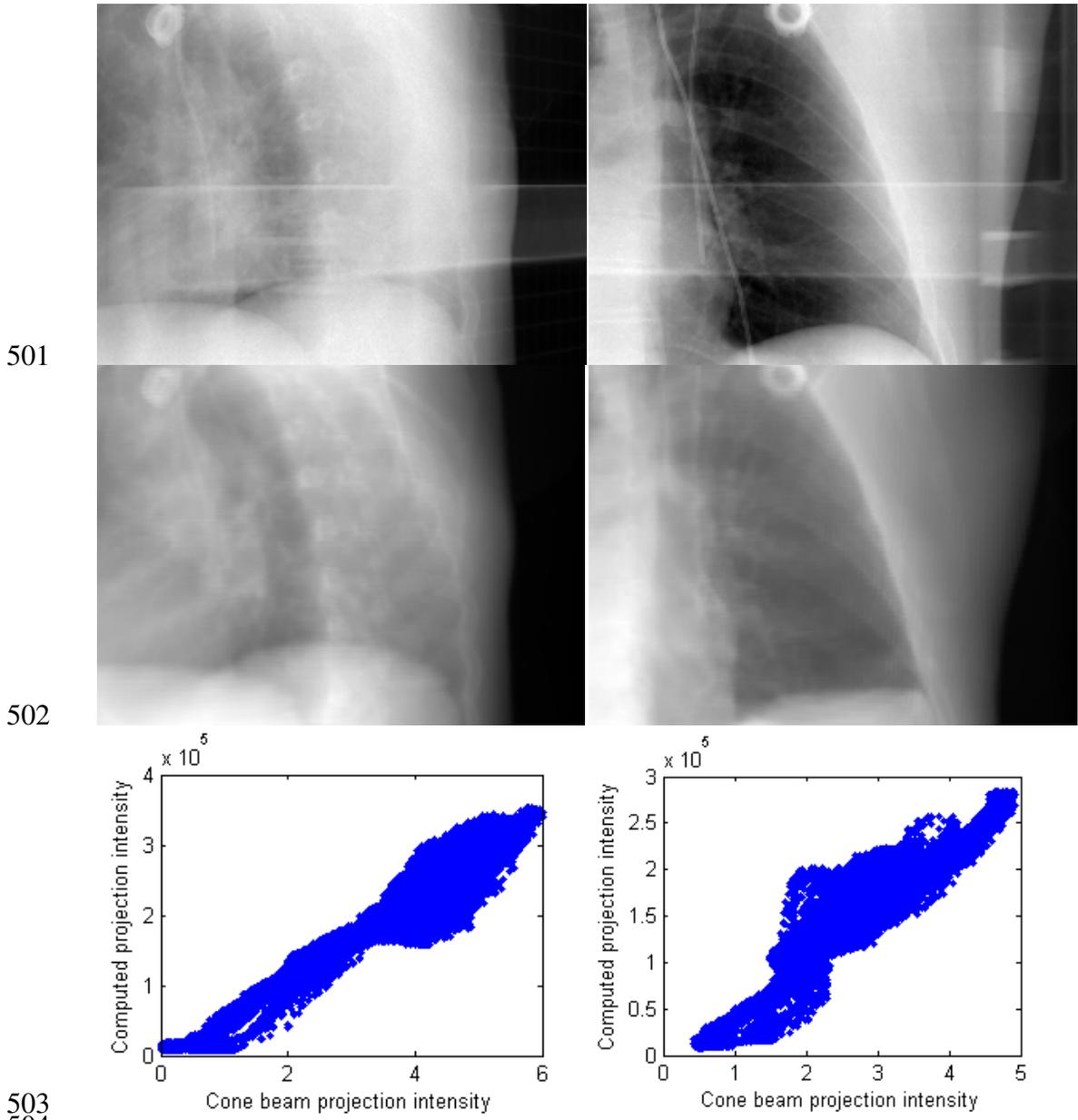

**Fig. 7.** Measured cone beam projections after preprocessing of the image intensities (logarithm transform followed by reversion of the sign) (*top row*), computed projections of the reconstructed image (*middle row*), and scatter plots between the intensities of the above two images (*bottom row*) for patient 1. Left column corresponds to the EOE phase in Fig. 4 (arrow "a"); right column corresponds to the EOI phase in Fig. 4 (arrow "b"). A linear model between the two image intensities is able to explain 94% and 91% of the variance in the left and right subplot, respectively. Both linear models are significant ($p < 0.001$).

In this work, we have assumed a linear relationship between the image intensities of the cone beam projection and of the computed projection of the reconstructed image, in order to account for the differences between the two image "modalities". One may wonder if such an assumption is accurate in reality. After all, the x-ray energies as well as their spectra used for 4DCT and CBCT scans may well be different. There are several physical processes in reality (*e.g.*, scattering) which are not simulated in our computation of simulated x-ray projection. Figure 7 shows the scatter plots between the image intensities of the preprocessed cone beam projection and of the computed projection of the reconstructed image at two different projection angles for patient 1, corresponding to the two projections shown in Fig. 5. It seems that a linear model is sufficient in both cases: more than 90% of the variance can be explained by a simple linear model. However, in the right subplot, because of more apparent interference from the treatment couch in the cone beam projection (which is absent in the computed projections), especially near the right edge, the linear model is less accurate than in the left subplot. This effect can be mitigated by pre-scanning the treatment couch and adding it to the reference image so that the couch will be also present in the computed projections.

This work has focused on x-ray projections with rotational geometry. The same principle can be easily applied to those with fixed-angle geometry, such as fluoroscopy. The only difference is that the projection matrix will be constant for fixed-angle geometry instead of varying at each angle for the rotational geometry considered in this work.

It is possible that during the course of the treatment patients may undergo anatomical changes on an inter-fractional basis, which could make the results worse if the lung motion model is built based on 4DCT acquired during patient simulation. In this case, a PCA lung motion model built from 4DCBCT during patient setup may overcome this difficulty. Another potential issue is when the tumor gets close to the heart. Then the tumor motion will be affected not only by breathing motion, but by heart motion too, which is not represented by the PCA motion model. This is a fundamental limitation of 4DCT, which is usually synchronized to respiration, not to cardiac motion.

Although the results of five lung cancer patients confirmed the validity of our algorithm on a preliminary basis, a more comprehensive study on a larger patient population is warranted. It is worth mentioning that the accuracy of the tumor localization and image reconstruction is directly affected by the PCA lung motion model, which is in turn influenced by the quality of the training data (4DCT or 4DCBCT). Therefore, it is important to minimize the motion artifacts in the training data, *e.g.*, by optimizing the scanning protocol or coaching the patients to breathe regularly whenever possible. A reliable deformable registration algorithm, especially one which can accurately model large motion (*e.g.*, around the diaphragm), is also likely to help improve the accuracy the PCA lung motion model. In this work, since there were no fiducial markers implanted in the lung cancer patients, clinician marked tumor positions had to be used to evaluate the tumor localization results. The accuracy of the ground truth is thus limited due to noisy projection images as well as relatively poor soft-tissue contrast. For validation purposes, it would be beneficial to have patient data with implanted fiducial markers, from which better ground truth can be derived.

## V.    Conclusions

In this work, we have presented an improved algorithm for 3D tumor localization from a single x-ray projection image by incorporating respiratory motion prediction. Through a comprehensive evaluation of our algorithm, we found the localization error to be less than 1 mm on average and around 2 mm at 95 percentile for both digital and physical phantoms, and within 2 mm on average and 4 mm at 95 percentile for five lung cancer patients on both axial and tangential directions. We also found that the localization accuracy is not affected by the breathing pattern, be it regular or irregular. The 3D localization can be achieved on a sub-second scale on GPU, requiring approximately 0.1 - 0.3 s for each x-ray projection.

567

### Acknowledgment

This work is supported in part by Varian Master Research Agreement. We would like to thank Dr. Paul Segars for providing the source code to generate NCAT phantom and NVIDIA for providing GPU cards for this project. We would also like to thank the anonymous reviewers who gave valuable comments and suggestions and helped improve this paper.

568
569
570
571
572
573
574

### Appendix

575
576

In this appendix, we describe in detail the optimization of the cost function in Eq. (2) with respect to $\mathbf{w}, a, b$. The optimization algorithm alternates between the following 2 steps:

577
578

$$\text{step 1: } \left(a_{n+1}, b_{n+1}\right)^T = \left(\mathbf{Y}^T \mathbf{Y}\right)^{-1} \mathbf{Y}^T \mathbf{P} \mathbf{f}_n \tag{A.1}$$

579

$$\text{step 2: } \mathbf{w}_{n+1} = \mathbf{w}_n - \mu_n \cdot \frac{\partial J_n}{\partial \mathbf{w}_n} \left/ \left\| \frac{\partial J_n}{\partial \mathbf{w}_n} \right\|_2 \right. \tag{A.2}$$

580

$$\text{where, } \mathbf{Y} = \left[\mathbf{y}, \mathbf{1}\right], \text{ and } \frac{\partial J}{\partial \mathbf{w}} = \frac{\partial \mathbf{x}}{\partial \mathbf{w}} \cdot \frac{\partial \mathbf{f}}{\partial \mathbf{x}} \cdot \frac{\partial J}{\partial \mathbf{f}} = 2 \cdot \mathbf{U}^T \cdot \frac{\partial \mathbf{f}}{\partial \mathbf{x}} \cdot \mathbf{P}^T \cdot \left(\mathbf{P} \cdot \mathbf{f} - a \cdot \mathbf{y} - b \cdot \mathbf{1}\right).$$

581
582

It is easy to see that the above alternating algorithm consisting of steps 1 and 2 is guaranteed to converge. In step 1, the update for a, b is the unique minimizer of the cost function with fixed $\mathbf{w}$. Step 2 is a gradient descent method with variable $\mathbf{w}$ and fixed a, b, where the step size $\mu_n$ is found by Armijo's rule[22] for line search. Therefore, the cost function always decreases at each step. Note that the cost function is lower bounded by zero. The above alternating algorithm is guaranteed to converge for all practical purposes. The parameters selected for line search are: initial step size $\mu_0 = 2$ and $\alpha = 0.1, \beta = 0.5$ as defined by Armijo's rule. These values generally lead to fast convergence and were used for all experiments in this study. The algorithm stops whenever the step size for current iteration is sufficiently small (fixed at 0.1 in this paper). The reason is that each PCA coefficients has been preconditioned so that their standard deviation is 10 in the training set; thus a step size of 0.1 can be considered sufficiently small and do not alter the results significantly. The algorithm is also terminated when the maximum number of iterations is reached (fixed at 10 in this paper). In the case of largest deformation from the reference image, it takes around 8 or 9 iterations for the algorithm to 'converge', in the sense that the cost function does not change much after that. Therefore, a maximum of 10 iterations should be sufficient to get convergent results.

583
584
585
586
587
588
589
590
591
592
593
594
595
596
597
598

At each iteration, given the updated PCA coefficients in Eq. (A.2), the DVF is updated according to Eq. (2) and the reconstructed image $\mathbf{f}_{n+1}$ is obtained through trilinear interpolation. Accordingly, $\partial \mathbf{f} / \partial \mathbf{x}$ has to be consistent with the interpolation process in order to get the correct gradient. Next, we derive the expression for $\partial \mathbf{f} / \partial \mathbf{x}$. Let $\mathbf{f}_0(i, j, k)$ and $\mathbf{f}(i, j, k)$ be the reference image and reconstructed image at iteration $n$ indexed by the integer set $\{(i, j, k) | 1 \leq i \leq N_1, 1 \leq j \leq N_2, 1 \leq k \leq N_3\}$ within the bounding box of its volume, and $\mathbf{x}_1(i, j, k), \mathbf{x}_2(i, j, k), \mathbf{x}_3(i, j, k)$ be the deformation vector fields corresponding to the three canonical directions, namely, lateral, AP, and SI, respectively. They can be equivalently

599
600
601
602
603
604
605
606

607 reorganized into a vector whose length is the total number of voxels in the image. Then
608 $\partial \mathbf{f}/\partial \mathbf{x} = \left[\partial \mathbf{f}/\partial \mathbf{x}_1; \partial \mathbf{f}/\partial \mathbf{x}_2; \partial \mathbf{f}/\partial \mathbf{x}_3\right].$
609

610 Let's look at $\partial \mathbf{f}/\partial \mathbf{x}_1$. First notice that:

611 $\mathbf{f}(i,j,k) = \mathbf{f}_0\left(i + \mathbf{x}_1(i,j,k), j + \mathbf{x}_2(i,j,k), k + \mathbf{x}_3(i,j,k)\right)$  (A.3)

612 Then we have:

613 $\dfrac{\partial \mathbf{f}(i_1,j_1,k_1)}{\partial \mathbf{x}_1(i_2,j_2,k_2)} = \delta(i_1 - i_2)\delta(j_1 - j_2)\delta(k_1 - k_2) \cdot \dfrac{\partial \mathbf{f}(i_1,j_1,k_1)}{\partial \mathbf{x}_1(i_2,j_2,k_2)}.$  (A.4)

614 This means that the Jacobian matrix $\partial \mathbf{f}/\partial \mathbf{x}_1$ is a diagonal matrix.

615

616 If the continuous patient geometry can be approximated by trilinear interpolation of the discrete
617 volumetric image, then we have:

618
$$\begin{aligned}
\mathbf{f}(i,j,k) = &\ \mathbf{f}_0\left(l,m,n\right)(1-z_1)(1-z_2)(1-z_3) + \mathbf{f}_0\left(l+1,m,n\right)z_1(1-z_2)(1-z_3) \\
&+ \mathbf{f}_0\left(l,m+1,n\right)(1-z_1)z_2(1-z_3) + \mathbf{f}_0\left(l,m,n+1\right)(1-z_1)(1-z_2)z_3 \\
&+ \mathbf{f}_0\left(l+1,m+1,n\right)z_1z_2(1-z_3) + \mathbf{f}_0\left(l+1,m,n+1\right)z_1(1-z_2)z_3 \\
&+ \mathbf{f}_0\left(l,m+1,n+1\right)(1-z_1)z_2z_3 + \mathbf{f}_0\left(l+1,m+1,n+1\right)z_1z_2z_3
\end{aligned}$$
(A.5)

619 where,

620 $l = \left\lfloor i + \mathbf{x}_1(i,j,k)\right\rfloor, m = \left\lfloor j + \mathbf{x}_2(i,j,k)\right\rfloor, n = \left\lfloor k + \mathbf{x}_3(i,j,k)\right\rfloor$ are the integer parts of the DVF

621 and $z_q = \left\{\mathbf{x}_q(i,j,k)\right\}, q = 1,2,3$ are the fractional parts of the DVF.

622

623 Notice that:

624 $\partial l/\partial \mathbf{x}_1 = \partial m/\partial \mathbf{x}_1 = \partial n/\partial \mathbf{x}_1 = 0$  (A.6)

625 and $\partial z_1/\partial \mathbf{x}_1 = 1,\ \ \partial z_2/\partial \mathbf{x}_1 = \partial z_3/\partial \mathbf{x}_1 = 0$  (A.7)

626 where, for simplicity, we have omitted the index $(i,j,k)$ for $\mathbf{x}_1$.

627

628 From Eq. (A.5)-(A.7), we immediately get:

629
$$\begin{aligned}
\partial \mathbf{f}(i,j,k)/\partial \mathbf{x}_1(i,j,k) = &\ \left[\mathbf{f}_0\left(l+1,m,n\right) - \mathbf{f}_0\left(l,m,n\right)\right](1-z_2)(1-z_3) \\
&+ \left[\mathbf{f}_0\left(l+1,m+1,n\right) - \mathbf{f}_0\left(l,m+1,n\right)\right]z_2(1-z_3) \\
&+ \left[\mathbf{f}_0\left(l+1,m,n+1\right) - \mathbf{f}_0\left(l,m,n+1\right)\right](1-z_2)z_3 \\
&+ \left[\mathbf{f}_0\left(l+1,m+1,n+1\right) - \mathbf{f}_0\left(l,m+1,n+1\right)\right]z_2z_3
\end{aligned}$$
(A.8)

630

631 Similar expressions can be derived for $\partial \mathbf{f}/\partial \mathbf{x}_2$ and $\partial \mathbf{f}/\partial \mathbf{x}_3$.

632

633 A few comments are appropriate at this point. Equation (A.8) is the exact derivative given the
634 continuous patient geometry obtained from trilinear interpolation in Eq. (A.5). Of course, any
635 reasonable kind of interpolation can be used here, *e.g.*, tricubic interpolation. Then the Jacobian
636 matrix $\partial \mathbf{f}/\partial \mathbf{x}_1$ does not have a nice diagonal form anymore, but may be block diagonal. It is clear
637 from Eq. (A.8) that $\partial \mathbf{f}/\partial \mathbf{x}$ is a linear combination of the spatial gradients of the reference image
638 evaluated at the neighboring eight grid points, weighted by the appropriate fractional parts of the
639 DVF. At this stage, $\partial \mathbf{f}/\partial \mathbf{x}$ contains only local information at the level of individual voxels. It is

640 the eigenvectors that effectively combine all the local information into global information and
641 lead to correct update for the PCA coefficients.
642
643 For trilinear interpolation, $\mathbf{f}$ is continuous but not differentiable at the boundaries of the voxel
644 grids, so that Eq. (A.6)-(A.7) do not hold in this situation. However, this happens with a
645 probability of 0. If it does happen, Eq. (A.8) simply computes the right-side derivative and will
646 not create numerical instability. What's more, even if it happens that the DVFs are integers at a
647 few voxels, the pooling effects of the eigenvectors will almost eliminates its influence by a vast
648 majority of other voxels.
649
650

## References

1. P. J. Keall, G. S. Mageras, J. M. Balter, R. S. Emery, K. M. Forster, S. B. Jiang, J. M. Kapatoes, D. A. Low, M. J. Murphy, B. R. Murray, C. R. Ramsey, M. B. Van Herk, S. S. Vedam, J. W. Wong and E. Yorke, "The management of respiratory motion in radiation oncology report of AAPM Task Group 76," Med Phys **33**, 3874-3900 (2006).

2. T. Harada, H. Shirato, S. Ogura, S. Oizumi, K. Yamazaki, S. Shimizu, R. Onimaru, K. Miyasaka, M. Nishimura and H. Dosaka-Akita, "Real-time tumor-tracking radiation therapy for lung carcinoma by the aid of insertion of a gold marker using bronchofiberscopy," Cancer **95**, 1720-1727. (2002).

3. F. Laurent, V. Latrabe, B. Vergier, M. Montaudon, J. M. Vernejoux and J. Dubrez, "CT-guided transthoracic needle biopsy of pulmonary nodules smaller than 20 mm: results with an automated 20-gauge coaxial cutting needle," Clin Radiol **55**, 281-287 (2000).

4. J. D. Hoisak, K. E. Sixel, R. Tirona, P. C. Cheung and J. P. Pignol, "Correlation of lung tumor motion with external surrogate indicators of respiration," Int J Radiat Oncol Biol Phys **60**, 1298-1306 (2004).

5. Y. Tsunashima, T. Sakae, Y. Shioyama, K. Kagei, T. Terunuma, A. Nohtomi and Y. Akine, "Correlation between the respiratory waveform measured using a respiratory sensor and 3D tumor motion in gated radiotherapy," Int J Radiat Oncol Biol Phys **60**, 951-958 (2004).

6. Y. Cui, J. G. Dy, G. C. Sharp, B. Alexander and S. B. Jiang, "Multiple template-based fluoroscopic tracking of lung tumor mass without implanted fiducial markers," Phys Med Biol **52**, 6229-6242 (2007).

7. Q. Xu, R. J. Hamilton, R. A. Schowengerdt, B. Alexander and S. B. Jiang, "Lung Tumor Tracking in Fluoroscopic Video Based on Optical Flow," Medical Physics **35**, 5351-5359 (2008).

8. T. Lin, L. I. Cervino, X. Tang, N. Vasconcelos and S. B. Jiang, "Fluoroscopic tumor tracking for image-guided lung cancer radiotherapy," Phys Med Biol **54**, 981-992 (2009).

9. J. H. Lewis, R. Li, W. T. Watkins, J. D. Lawson, W. P. Segars, L. I. Cervino, W. Y. Song and S. B. Jiang, "Markerless lung tumor tracking and trajectory reconstruction using rotational cone-beam projections: a feasibility study," Phys Med Biol **55**, 2505-2522 (2010).

10. R. Zeng, J. A. Fessler and J. M. Balter, "Estimating 3-D respiratory motion from orbiting views by tomographic image registration," IEEE Trans Med Imaging **26**, 153-163 (2007).

11. Q. Zhang, A. Pevsner, A. Hertanto, Y. C. Hu, K. E. Rosenzweig, C. C. Ling and G. S. Mageras, "A patient-specific respiratory model of anatomical motion for radiation treatment planning," Med Phys **34**, 4772-4781 (2007).

12. D. A. Low, P. J. Parikh, W. Lu, J. F. Dempsey, S. H. Wahab, J. P. Hubenschmidt, M. M. Nystrom, M. Handoko and J. D. Bradley, "Novel breathing motion model for radiotherapy," Int J Radiat Oncol Biol Phys **63**, 921-929 (2005).

13. R. Li, J. H. Lewis, X. Jia, T. Zhao, W. Liu, S. Wuenschel, J. Lamb, D. Yang, D. A. Low and S. B. Jiang, "A PCA-based lung motion model," Phys Med Biol *submitted* (2010).

14. R. Li, X. Jia, J. H. Lewis, X. Gu, M. Folkerts, C. Men and S. B. Jiang, "Real-time volumetric image reconstruction and 3D tumor localization based on a single x-ray projection image for lung cancer radiotherapy," Med. Phys. **37**, 2822-2826 (2010).

15. M. J. Murphy, J. Jalden and M. Isaksson, presented at the Proc. 16th Int. Conf. on Computer Assisted Radiology (CARS 2002), Paris, France, 2002 (unpublished).

16. G. C. Sharp, S. B. Jiang, S. Shimizu and H. Shirato, "Prediction of respiratory tumour motion for real-time image-guided radiotherapy," Phys Med Biol **49**, 425-440 (2004).

17. D. Ruan, J. A. Fessler and J. M. Balter, "Real-time prediction of respiratory motion based on local regression methods," Phys Med Biol **52**, 7137-7152 (2007).

18. M. Chen, W. Lu, Q. Chen, K. J. Ruchala and G. H. Olivera, "A simple fixed-point approach to invert a deformation field," Med Phys **35**, 81-88 (2008).

19. X. Gu, H. Pan, Y. Liang, R. Castillo, D. Yang, D. Choi, E. Castillo, A. Majumdar, T. Guerrero and S. B. Jiang, "Implementation and evaluation of various demons deformable image registration algorithms on a GPU," Phys Med Biol **55**, 207-219 (2010).

20. W. P. Segars, "Development and application of the new dynamic NURBS-based cardiac-torso (NCAT) phantom," Ph.D. Dissertation (2001).

21. R. L. Siddon, "Fast calculation of the exact radiological path for a three-dimensional CT array," Medical Physics **12**, 252-255 (1985).

22. M. S. Bazaraa, H. D. Sherali and C. M. Shetty, *Nonlinear Programming: Theory and Algorithms*. (Hoboken: A John Wiley & Sons, 2006).

[a] Email: sbjiang@ucsd.edu